\documentclass[%
 reprint,
 amsmath,amssymb,
 aps,
prb,
]{revtex4-1}

\usepackage{textcomp}
\usepackage{graphicx}% Include figure files
\usepackage{dcolumn}% Align table columns on decimal point
\usepackage{bm}% bold math
\usepackage{tabularx} % Ruled Tabular
\usepackage{lipsum}
\usepackage{multirow}
\usepackage{times}

\def\asih{{\it a}-Si:H}
\def\asi{{\it a}-Si}
\def\ac{{\it a}-C}

\begin{document}

\preprint{APS/123-QED}

\title{Large and realistic models of Amorphous Silicon }% Force line breaks with \\
%\thanks{}%

\author{Dale Igram}
\email{di994313@ohio.edu}
\author{Bishal Bhattarai}
\email{bb248213@ohio.edu}
\affiliation{Department of Physics and Astronomy \\ 
Condensed Matter and Surface Science Program (CMSS) \\
 Ohio University, Athens, Ohio 45701,USA}%Lines break automatically or can be forced with \\
\author{Parthapratim Biswas }%
\email{partha.biswas@usm.edu}
\affiliation{Department of Physics and Astronomy \\
The University of Southern Mississippi Hattiesburg, Mississippi 39406, USA}%
\author{D. A. Drabold }%
\email{drabold@ohio.edu}
\affiliation{Department of Physics and Astronomy \\
Nanoscale and Quantum Phenomena Institute (NQPI) \\
Ohio University, Athens, Ohio 45701, USA}%

\date{\today}% It is always \today, today,

\begin{abstract}
Amorphous silicon ({\asi}) models are analyzed for structural, electronic and vibrational characteristics.
Several models of various sizes have been computationally fabricated for this analysis.
It is shown that a recently developed structural modeling algorithm known as force-enhanced atomic refinement (FEAR) provides results
in agreement with experimental neutron and x-ray diffraction data while producing a total energy below conventional schemes. We also show that a large model
($\sim 500$ atoms) and a complete basis is necessary to properly describe vibrational and thermal properties. We compute the density for \asi, 
and compare with experimental results.

\end{abstract}

\pacs{Valid PACS appear here}% PACS, the Physics and Astronomy
\maketitle

%\tableofcontents

\section{\label{sec:level1}INTRODUCTION }

Amorphous silicon ({\asi}) and its hydrogenated counterpart ({\asih}) 
continue to play an important role in technological applications, such 
as thin-film transistors, active-matrix displays, image-sensor arrays, 
multi-junction solar cells, multilayer color detectors, thin-film 
position detectors, etc.~\cite{Street1} 
While a number of traditional methods, based on Monte Carlo and molecular-dynamics 
simulations, were developed in the past decades by directly employing 
classical or quantum-mechanical force fields -- from 
the event-based Wooten-Winer-Weaire (WWW)~\cite{WWW1,WWW2} bond-switching 
algorithm and the activation-relaxation technique (ART)~\cite{Barkema1, Barkema2}
to the conventional melt-quench (MQ) molecular-dynamics 
simulations~\cite{DraboldEuro1,Tersoff1,NAMarks1,car,dave,cooper2000} -- none 
of the methods utilize prior knowledge or experimental information in the 
simulation of atomistic models of complex materials. 
It is now widely accepted that dynamical methods perform rather poorly 
to generate high-quality (i.e., defect-free) continuous-random-network 
(CRN) models of amorphous silicon by producing too many coordination 
defects (e.g., 3- and 5-fold coordinated atoms) in the networks.  While the 
WWW algorithm and the ART can satisfactorily 
address this problem by producing 100\% defect-free CRN models 
of {\asi}, a direct generalization of the WWW 
algorithm for multicomponent systems is highly nontrivial in the absence of 
sufficient information on the bonding environment of the atoms.  
Likewise, the ART requires a detailed knowledge of the local minima 
and the saddle points on a given potential-energy surface in order 
to determine suitable low-lying minima that correspond to defect-free 
CRN models of amorphous silicon. 
On the other hand, the availability of high-precision experimental 
data from diffraction, infrared (IR), and nuclear magnetic 
resonance (NMR) measurements provide unique opportunities to 
develop methods, based on information paradigm, where one 
can directly incorporate experimental data in 
simulation methodologies. The reverse Monte Carlo (RMC) method~\cite{McGreevy1,
McGreevy2,McGreevy3,Biswas_rmc_2004} is an 
archetypal example of this approach, where one attempts to determine 
the structure of complex disordered/amorphous solids by inverting 
experimental diffraction data. Despite its simplicity and elegance, the 
method produces unphysical structures using diffraction data 
only.  While inclusion of appropriate geometrical/structural constraints 
can ameliorate the problem, the generation of high-quality 
models of {\asi}, using constrained RMC simulations, has 
been proved to be a rather difficult optimization problem and 
satisfactory RMC models of {\asi} have not been reported in 
the literature to our knowledge. 
The difficulty associated with the inversion of diffraction data 
using RMC simulations has led to the development of a number of hybrid 
approaches in the past decade.\cite{Opletal2,Cliffe1}  Hybrid approaches retain the 
spirit of the RMC philosophy as far as the use of experimental data 
in simulations is concerned but go beyond RMC by using an 
extended penalty function, which involves total energy and 
forces from appropriate classical/quantum-mechanical force 
fields, in addition to few structural or geometrical constraints. 
The experimentally constrained molecular relaxation~\cite{ecmr1,ecmr2} (ECMR), 
the first-principle assisted structural solutions~\cite{FPASS} (FPASS), 
and the recently developed force-enhanced atomic 
relaxation~\cite{Anup1,Anup2,Anup3,Bhattarai3} (FEAR) 
are a few examples of hybrid approaches, which have successfully 
incorporated experimental information in atomistic simulations 
to determine structures consistent with both theory and 
experiments. Recently, the FEAR has been applied successfully 
to simulate amorphous carbon ({\ac}).~\cite{Bhattarai3} This is particularly 
notable as the latter can exist in a variety of complex carbon bonding 
environment, which makes it very difficult to produce {\ac} from {\it 
ab initio} molecular-dynamics simulations due to the lack of {\it glassy} 
behavior and the WWW bond-switching algorithm in the absence of prior knowledge of the 
bonding states of C atoms in {\ac} (e.g., the ratio of $sp^2$- versus $sp^3$-bonded 
C atoms with a varying mass density). In this paper, we show that the information-based 
FEAR approach can be employed effectively to large-scale simulations 
of {\asi} consisting of 1000 atoms. The resulting models have been 
found to exhibit superior structural, electronic, and vibrational 
properties of {\asi} as far as the existing RMC and {\it ab initio} 
MD models are concerned in the literature.

The rest of paper is as follows. In section II, we discuss 
the computational methodology associated with the generation of 
CRN models using the FEAR method. This is followed by the 
validating properties of FEAR models with particular 
emphasis on the structural, electronic, vibrational, and thermal 
properties in section III. Section IV presents the conclusions 
of our work.

\begin{figure*}[!ht]
\begin{minipage}[b]{.475\textwidth}
  \centering
  \includegraphics[width=0.95\linewidth]{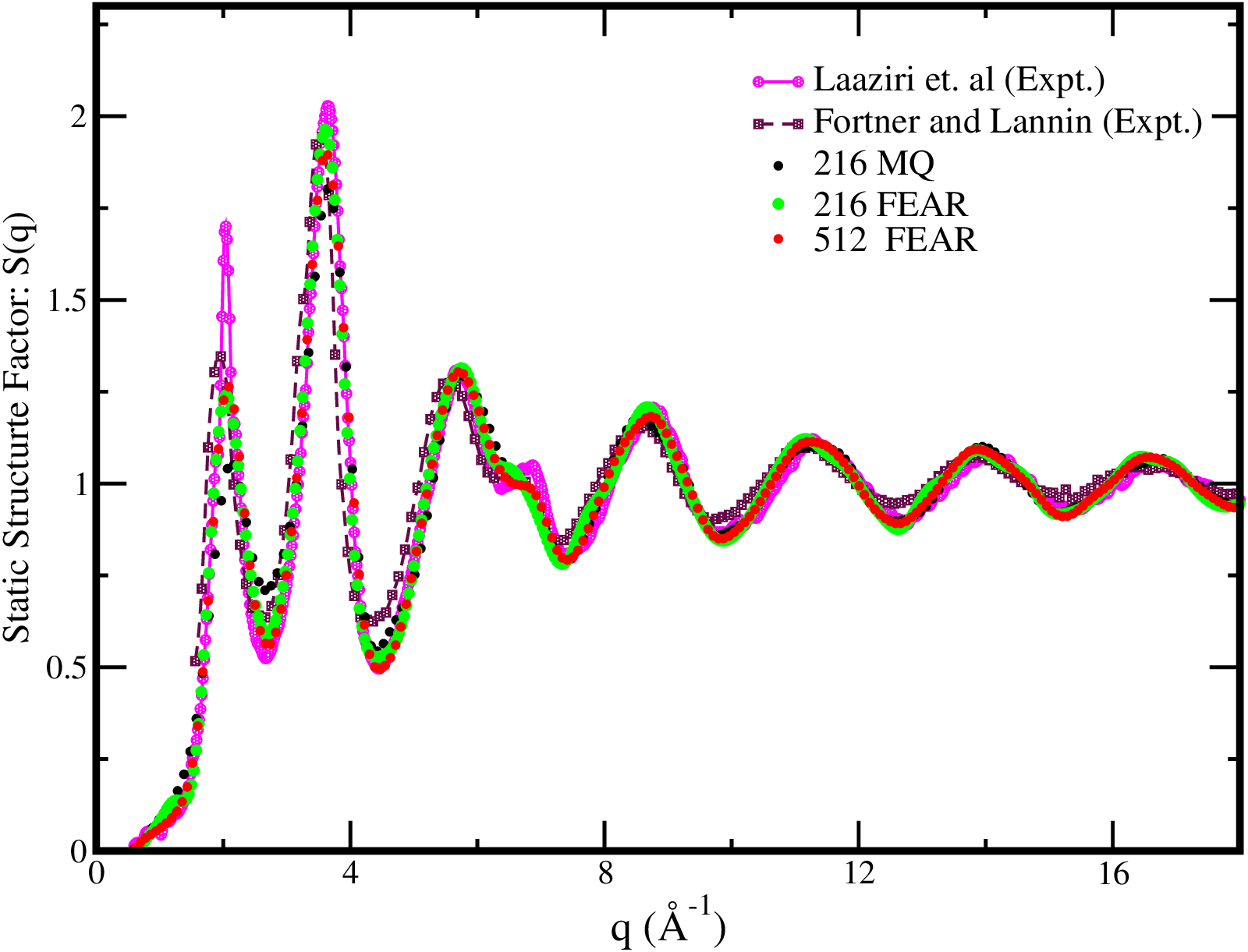}
\end{minipage}\hspace{0.2cm}
\begin{minipage}[b]{.475\textwidth}
  \centering
  \includegraphics[width=0.95\linewidth]{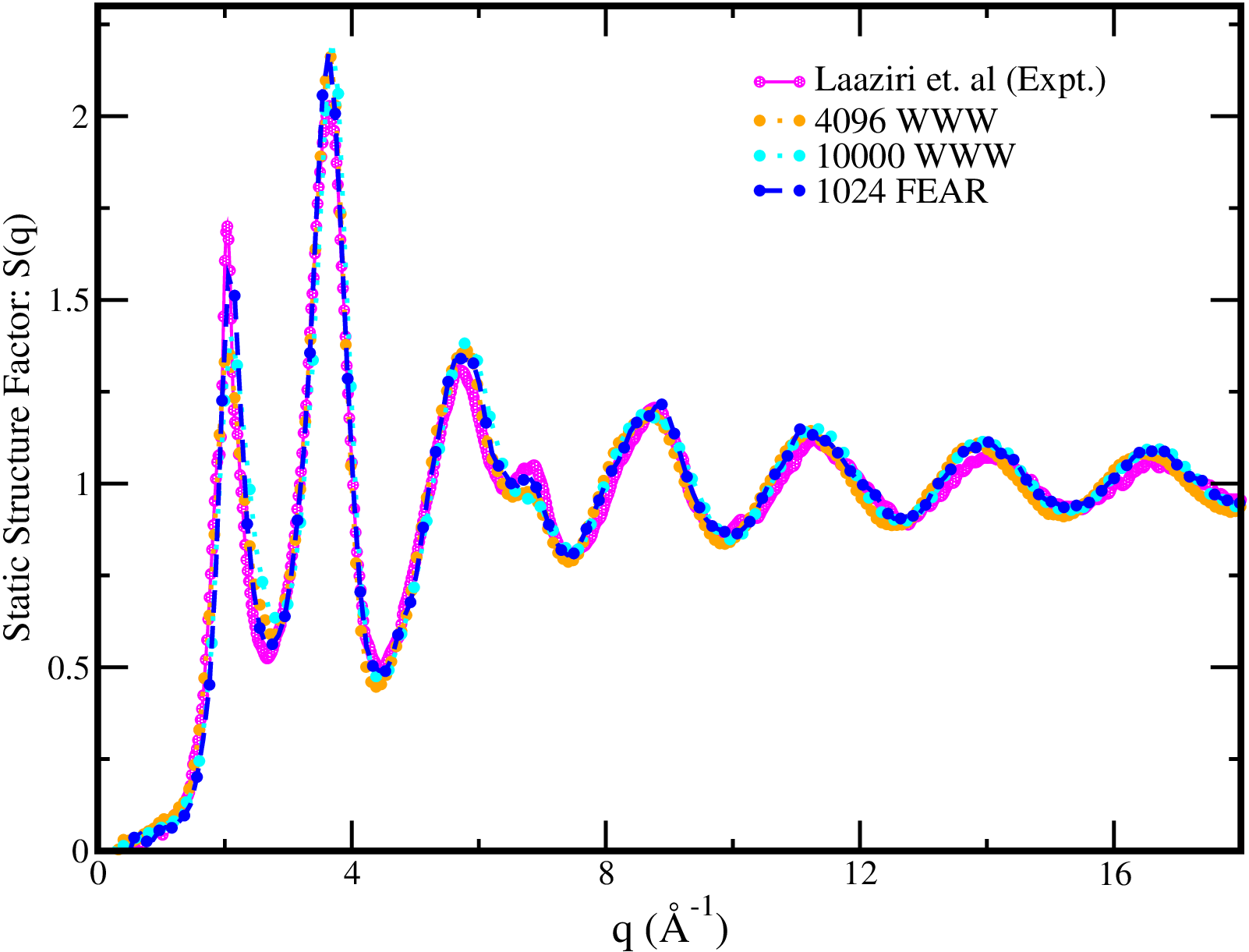}
\end{minipage}\vspace{0.1cm}

\caption{(Color online)  Structure factor for different models and their comparison with experiments.\cite{Laaziri1,Fortner1}}
\label{fig:fig1023111}
\end{figure*}

\begin{figure*}[!ht]
\begin{minipage}[b]{.475\textwidth}
  \centering
  \includegraphics[width=0.95\linewidth]{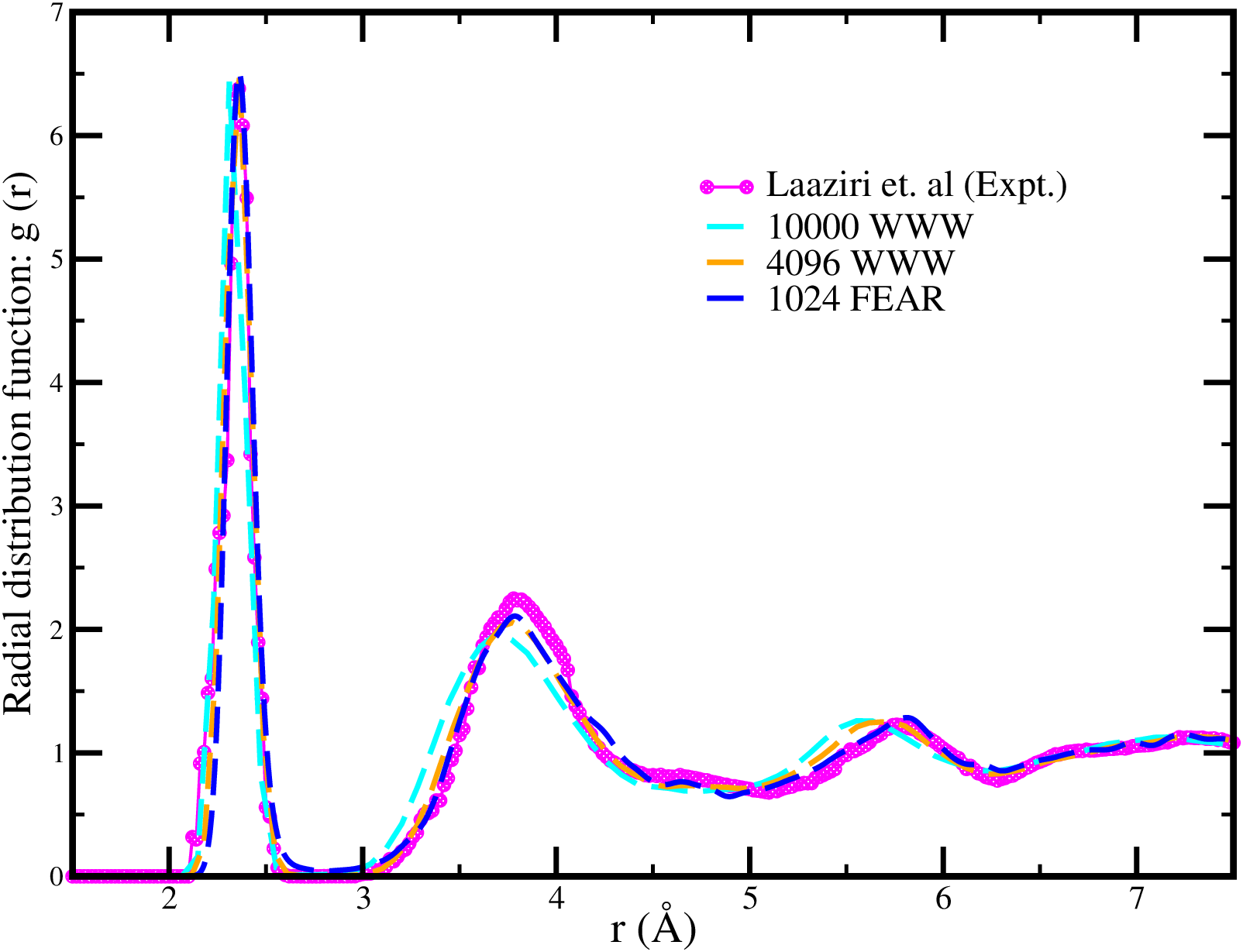}
\end{minipage}\hspace{0.2cm}
\begin{minipage}[b]{.475\textwidth}
  \centering
  \includegraphics[width=1.025\linewidth]{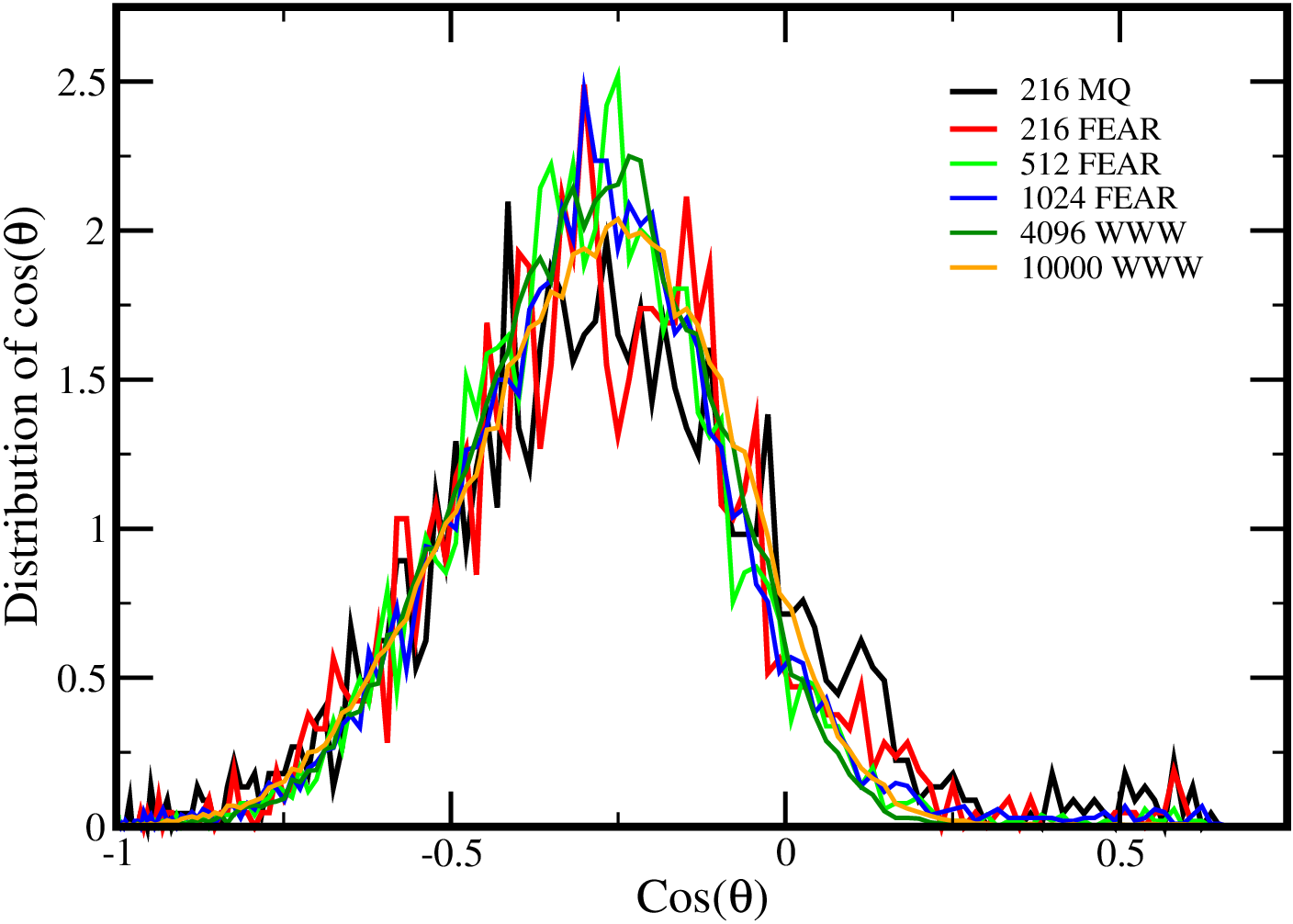}
\end{minipage}\vspace{0.1cm}

\caption{(Color online) (left panel)  Radial distribution function of different models and their comparison with experiment\cite{Laaziri1}, (right panel) Plot of bond-angle distribution for the six models.}
  
\label{fig:fig1023222}
\end{figure*}

\begin{table*}[!ht]
\centering
\caption{\label{tab:table3}Nomenclature and details of our models: Length of the cubic box(L), position of first ($r_1$) and second ($r_2$) peak of RDF, Average coordination number (n), percentage of 
3-fold, 4-fold and 5-fold coordinated atoms, Free Energy per atom of the final VASP relaxed models($E_0$).}
%\begin{ruledtabular}
\vspace{0.3cm}
\begin{tabular}{cccccccccc}
\hline\hline
\rule{0pt}{4mm}
 %&\multicolumn{2}{c}{$D_{4h}^1$}&\multicolumn{2}{c}{$D_{4h}^5$}\\
 %Ion&1st alternative&2nd alternative&lst alternative
 Model & $ L($\AA$)$ & $r_1($\AA$)$ & $r_2($\AA$)$ & n & 3-fold $\%$ & 4-fold $\%$ & 5-fold$\%$ & $E_0(eV/atom)$ \\[0.5ex]\\ \hline
\rule{0pt}{4mm}

216MQ & 16.28 & 2.36 & 3.81  & {4.083} & 0.93  & {87.03} &  11.57 & {0.000}  \\
\rule{0pt}{4 mm}

216FEAR & 16.28 & 2.36 & 3.81  & {4.028} & 1.39 & {94.44} & 4.17 & {-0.002}  \\
\rule{0pt}{4 mm} 

512FEAR & 21.71& 2.35 & 3.82  & {4.008} & 1.17  & {95.90} & 2.73 & {-0.044}  \\  % inserting body of the table
 \rule{0pt}{4 mm}
 
1024FEAR & 27.35 & 2.36 & 3.79 & {4.018} & 2.34  & {94.53} & 3.13 & {-0.035} \\
\rule{0pt}{4 mm}

4096WWW & 43.42 & 2.36 & 3.78 & {4.004} & 0.05 & {99.46} & 0.49 & {---}  \\
\rule{0pt}{4 mm}

10000WWW & 57.32 & 2.31 & 3.69 & {4.014} & 0.04  & {98.60} & 1.30 & {---}  \\
\hline\hline
\end{tabular}
\end{table*}

\section{\label{sec:level2} Methodology and Models}
For this study, three model sizes (216, 512 and 1024 atoms) were implemented with FEAR  and compared with experimental data. 
Several algorithms and codes were utilized for the preparation of the models; namely, FEAR\cite{Anup1,Anup2,Anup3}, RMCProfile\cite{Tucker2}, SIESTA\cite{Siesta1}
and VASP.\cite{Kresse2,Vaspgpu1,Vaspgpu2}

A random starting structure was constructed for each of the models and was refined by fitting to the experimental pair correlation 
functions $g(r)$ and/or the static structure factor $S(q)$ by employing RMCProfile. 
The refined structure is relaxed using conjugate gradient (CG) in SIESTA. 
The relaxed-refined structure is then refined by RMCProfile. This cyclic process is repeated until convergence is achieved. 
For completeness the converged structure is then fully relaxed by VASP (plane wave LDA). 

The partial refinement steps in RMCProfile were carried out with a minimum distance between atoms of 2.10 $\mathring{A}$ and maximum move 
distance of 0.15 $\mathring{A}$ -- 0.35 $\mathring{A}$.
The partial relaxation steps utilized SIESTA with a single-$\zeta$ basis set, Harris functional at constant volume, exchange-correlation functional with local-density approximation (LDA),
periodic boundary conditions and a single relaxation step. The final relaxation step employed VASP with a plane-wave basis set, plane-wave cutoff of $350-450$ eV, energy difference criteria of $10^{-4}-10^{-5}$. 
The fully relaxed calculations were performed for $\Gamma ({\vec{k}}=0)$.
For all the FEAR models, 
we have used structure factor data from \textit{Laaziri et.al.}\cite{Laaziri1} for RMC refinement.

The three FEAR models and 216 MQ model have a number density of about 0.05005 atom/$\mathring{A}^3$, which is associated with atomic density of 2.33 $g/cm^{-3}$ (for details Table I). The 216 MQ model was fabricated by taking a set of random coordinates and equilibrating these coordinates at 3000K for 6ps, 
followed by cooling from 3000K to 300K within 9 ps, then equilibration at 300K for 4.5 ps, and a full relaxation at 300K. The MQ calculations were performed with a step size of 1.5 fs.

We have also considered two large (4096 atom and 10,000 atom) WWW\cite{WWW1,WWW2} models in our comparison. These two WWW models were relaxed using SIESTA with a single-$\zeta$ basis set, LDA at 
constant volume utilizing Harris functional.\footnote{We minimized our 4096 WWW model to a have forces less than $0.01$ $eV/\mathring{A}$
and for the 10,000 WWW model after $\sim 100$ CG steps, RMS force of $0.024$ $eV/\mathring{A}$ was obtained.}

\section{\label{sec: level4} RESULTS AND DISCUSSION}

\subsection{\label{sec:level4a} Structural Properties}

A comparison of structure factors for the six models 216 MQ, 216 FEAR, 512 FEAR, 1024 FEAR, 4096 WWW and 10,000 WWW models with respect to 
experiment\cite{Laaziri1,Fortner1} is shown in Fig. 1. 
From, Fig. 1(left panel) we can clearly observe that these models of up to 512 atoms is insufficient to resolve the first peak occurring at low q.
In contrast, the 1024 FEAR model does well even in comparison to much larger models as seen in Fig. 1(right panel). This is also indicated in the real space 
information $g(r)$ (Fig. 2), where we observe that 10000 WWW model is slightly shifted as compared to the experiment\cite{Laaziri1} for the first
and second neighbors peak. We report the details of our simulation and important observables in Table I.

From Table I, we observe that there are some defects in our models. These structural defects arise due to a small fraction ($\sim 5\%$) of over co-ordinated and under co-ordinated atoms.
This explains the fact that all of our models have coordination value slightly above perfect four-fold coordination. Experimentally, it is also
observed that a-Si does not posses a perfect four fold coordination.\cite{Laaziri1,Fortner1} Our final models obtained after relaxation attain energies (eV/atom) equal or less than models obtained from MQ.

We further show our plots of bond-angle distribution in Fig. 2 (right panel) to attest 
accuracy of FEAR models. As seen in Fig. 2 the peak of the bond angle is close to the value of tetrahedral angle 109.47$^{o}$. Similarly, from ring statistics (Fig.3) 
we observe that these a-Si networks mostly prefer a ring size of 5,6,7. Small rings (mostly 3-membered rings) are responsible for a unrealistic peak seen in unconstrained RMC\cite{Anup1} 
at an angle around $\sim 60^{o}$. \textit{Opletal et. al.} have proposed use of a constraint for removal of these highly constrained 3 membered rings in 
several of their works.\cite{Opletal2,Opletal5} FEAR method which incorporates accurate \textit{ab initio} interaction enables us to remove these high energy
structures without satisfying an extra criterion.

\begin{figure}[!ht]
  \centering
   \includegraphics[width=5.0 cm,]{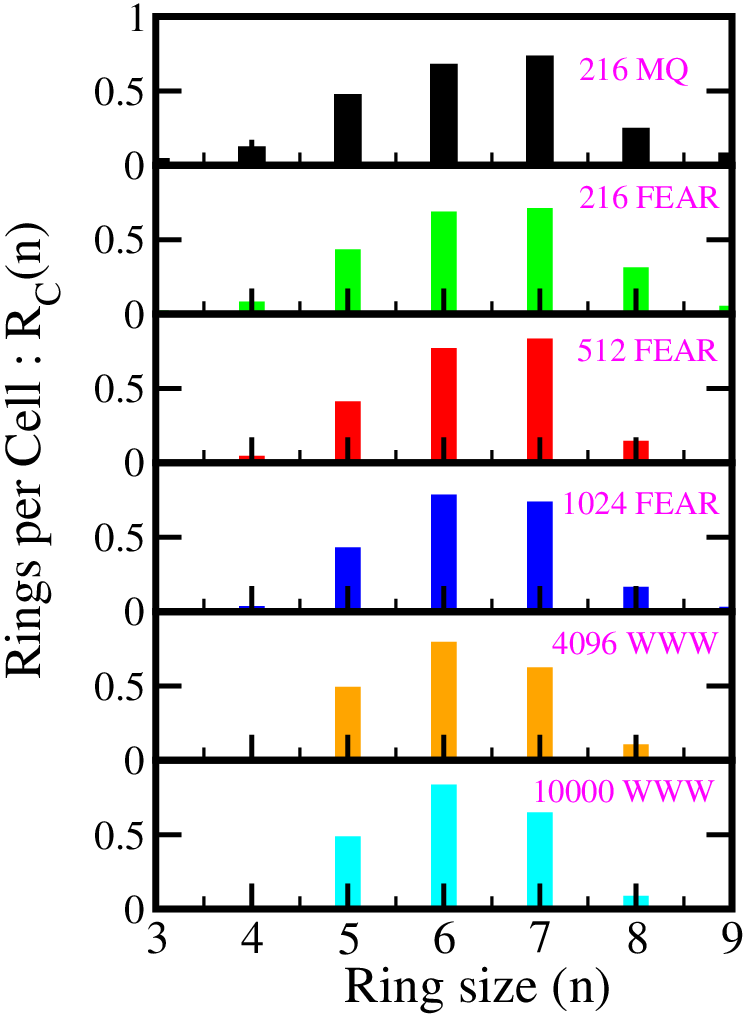}
  \caption{(Color online) Rings per cell ($R_C$)for the six models. The ring statistics were obtained using King's method\cite{King1} using ISAACS software\cite{Roux1}.}\label{40123}
\end{figure}

\subsection{\label{sec:level4b} Electronic Properties}

Electronic properties such as electronic density of states (EDOS) reveal crucial information regarding accuracy of models. In particular, \textit{Prasai et. al.} 
and others\cite{Kiran1,Los1} have used electronic information to aid in modeling amorphous system. Conversely, EDOS obtained for our models validate 
accuracy of our models. We have shown our plot of four models in Fig. 4. We have also studied the localization of electronic states by plotting 
inverse participation ratio (IPR) in 
conjunction with EDOS. We observe both plots with same qualitative resemblance with few localized states appearing near the Fermi energy ($E_F=0$). These 
localized states arise due to the defects in the model (3-fold and 5-fold atoms). 

\begin{figure}[!ht]
  \centering
   \includegraphics[width=9.0 cm,]{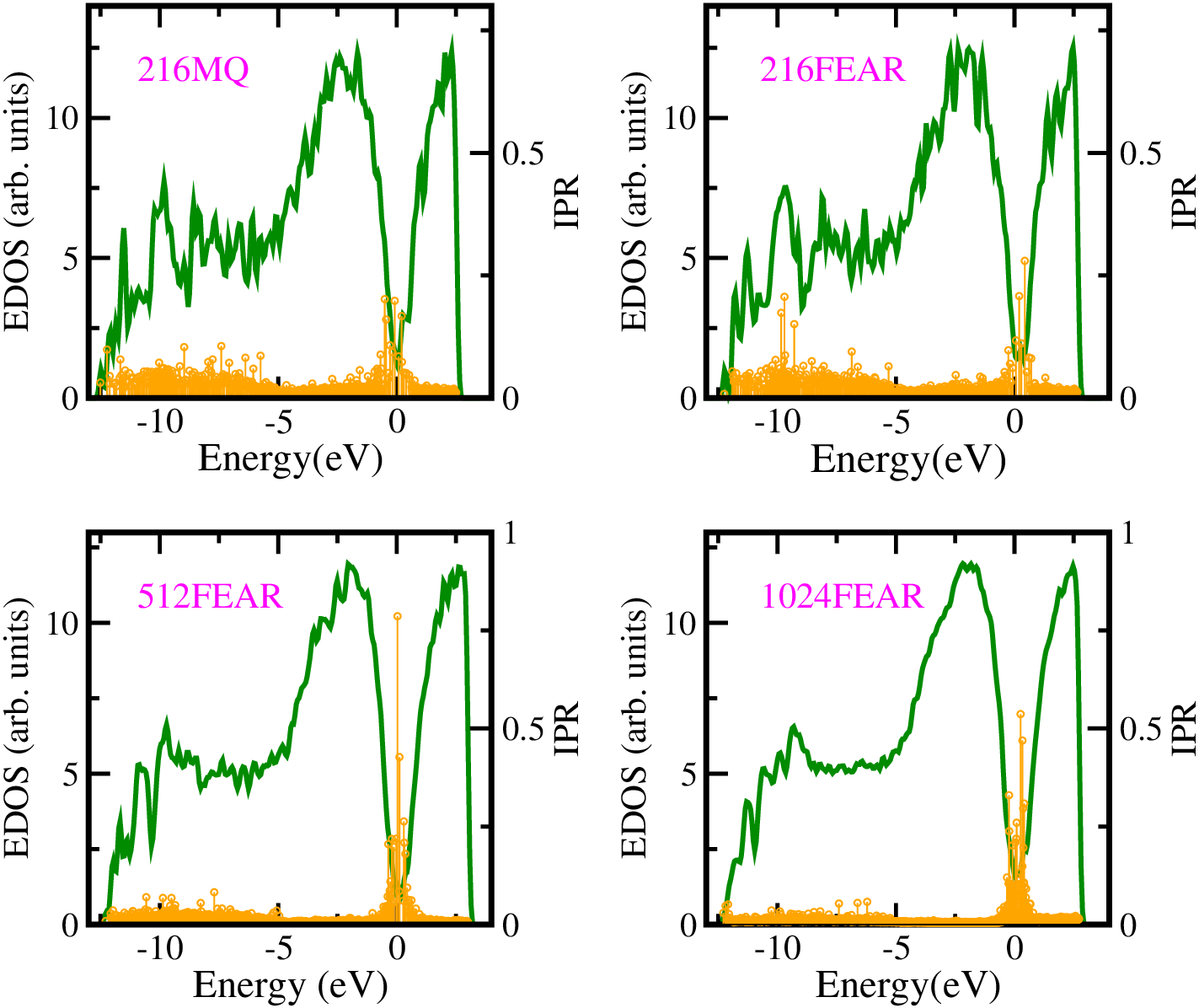}
  \caption{(Color online) Plot of Electronic density of states (EDOS($E_F$=0)) green-solid lines and Inverse participation ratio (IPR) yellow-drop lines.}\label{40132}
\end{figure}

We compare our large model of 4096 atoms along with our FEAR models. Due to the gigantic size of this model, we have used Harris Functional
and single-$\zeta$ basis set to evaluate the electronic density of states of these models. To our knowledge this is first time reporting
of an \textit{ab initio} based EDOS of {\asi}  models this big. \textit{Drabold et. al.} have previously carried out an extensive research regarding the exponential 
tail (valance and conduction) observed in amorphous silicon.\cite{DraboldUrbach2,DraboldUrbach1,DraboldUrbach3} We report our result of EDOS for these models in Fig. 5.
We observe that a 216 atom model gives us a very crude representation of these tails (valance and conduction). Meanwhile, FEAR models 512 and 1024 
compare well with the large WWW models. \textit{Fedders et. al}\cite{Fedders1} have revealed that 
the valance tail prefers short bonds while the conduction tail prefers long bonds.

\begin{figure}[!ht]
  \centering
   \includegraphics[width=5.5 cm,]{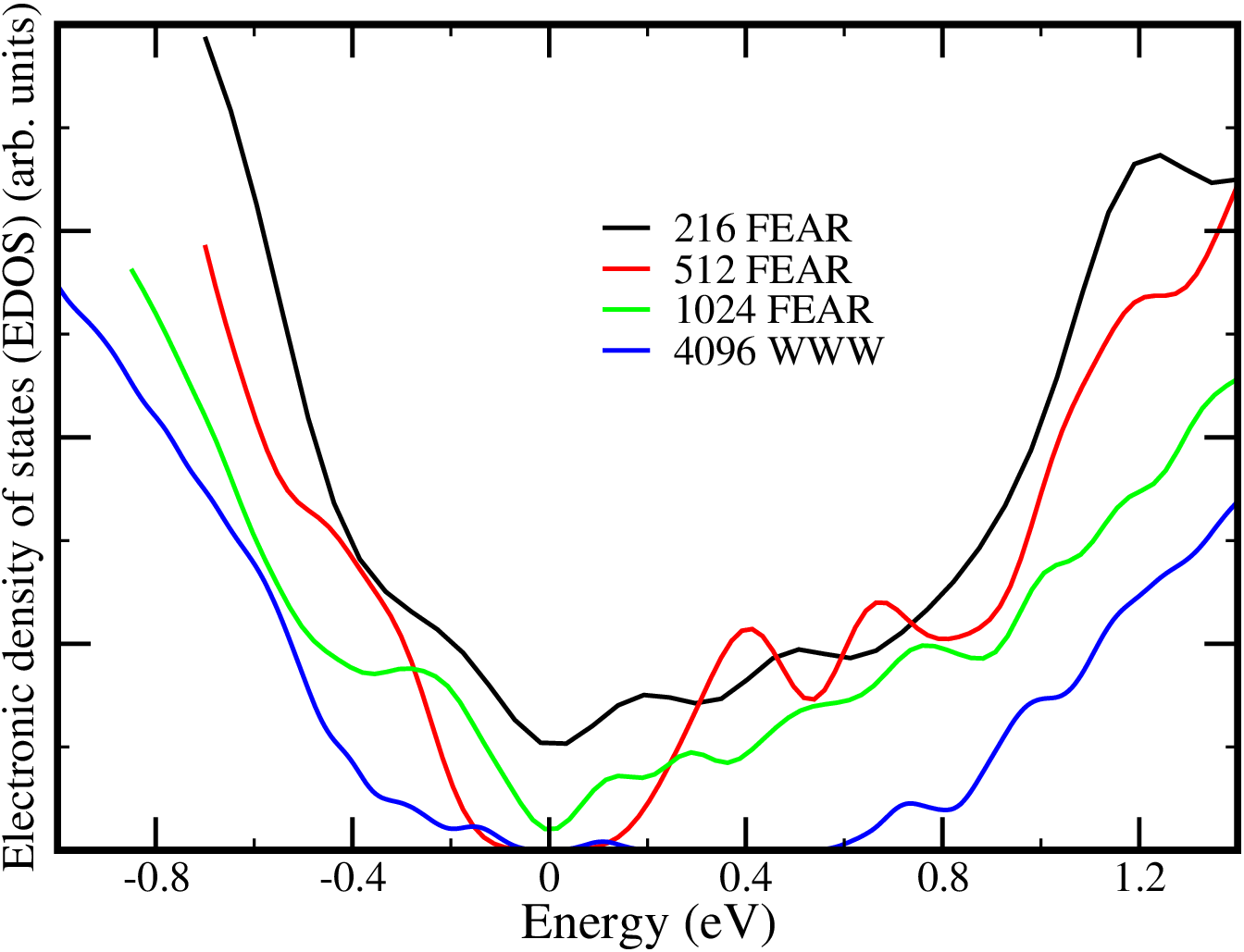}
  \caption{(Color online) Comparison of Electronic density of states ($E_F$=0) of different models obtained by SIESTA with single-$\zeta$ basis set with 
Harris functional.}\label{40111}
\end{figure}

\subsection{\label{sec:level5} Vibrational Density of States}

\begin{figure*}[!ht]
\begin{minipage}[b]{.475\textwidth}
  \centering
  \includegraphics[width=0.95\linewidth]{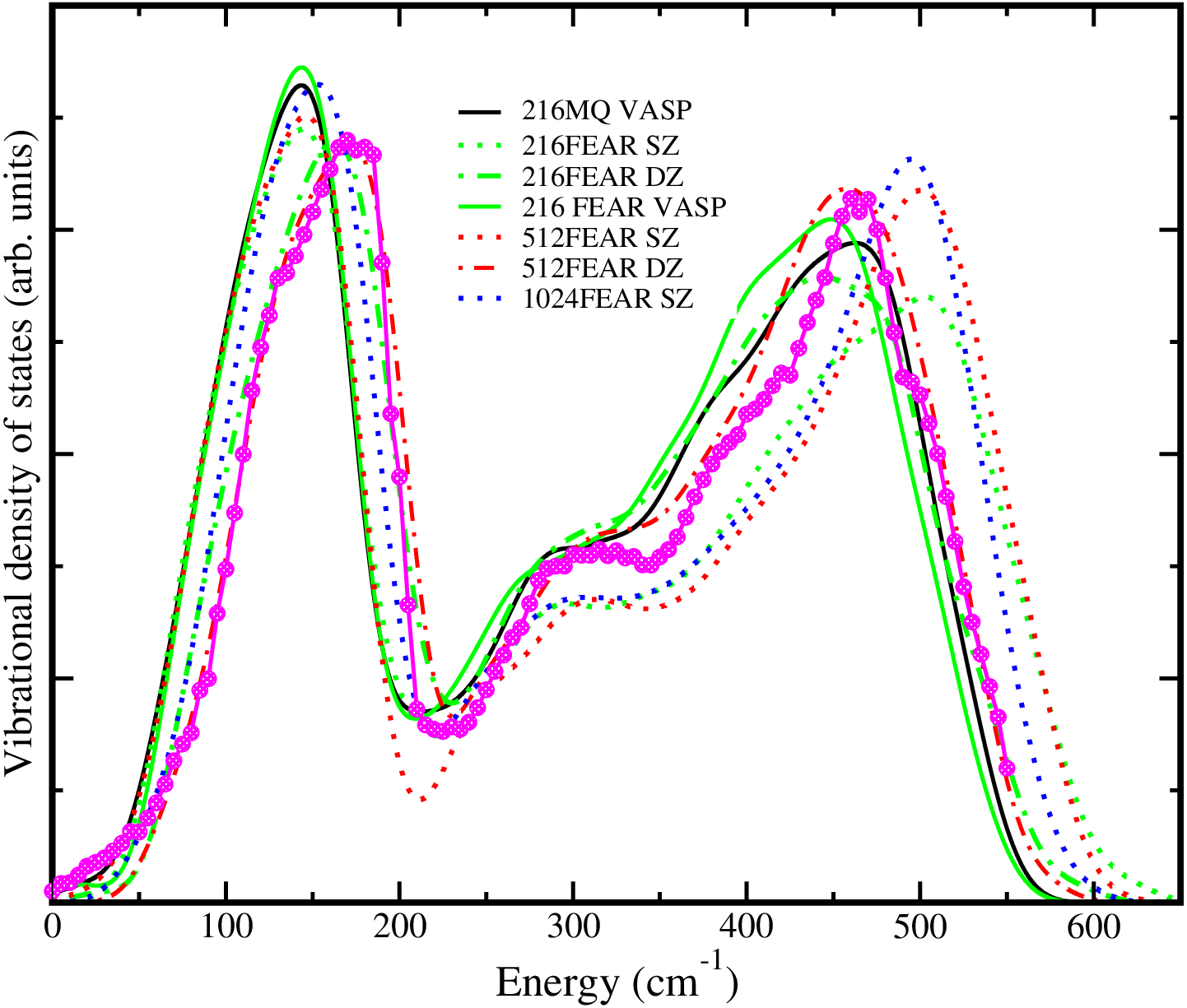}
\end{minipage}\hspace{0.2cm}
\begin{minipage}[b]{.475\textwidth}
  \centering
  \includegraphics[width=1.05\linewidth]{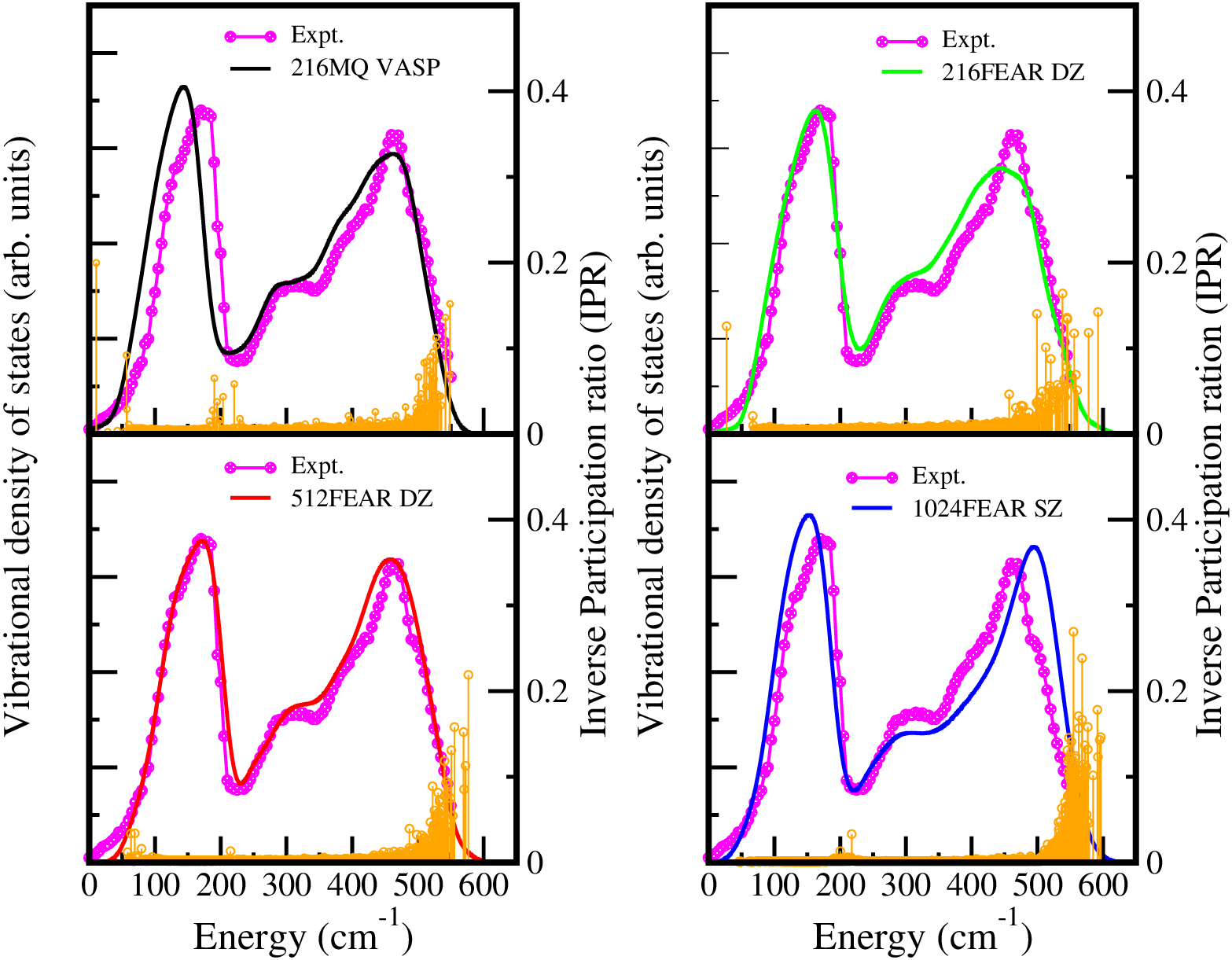}
\end{minipage}\vspace{0.1cm}

\caption{(Color online)  (left panel) Vibrational density of states (VDOS) obtained for different models using VASP-LDA, SIESTA-LDA(single-$\zeta$, SZ) and
SIESTA-LDA (double-$\zeta$, DZ), (right panel) Comparison of vibrational density of states (VDOS) with experimental results\cite{Kamitakahara1} (Note the almost perfect agreement for the 512 DZ calculation). The yellow
drop-lines shows Inverse participation ration (IPR), IPR measures localization of Eigen modes.}
  
\label{fig:fig1024444}
\end{figure*}

\subsubsection{Vibrational Properties}

Vibrational density of states (VDOS) provides key information about the local bonding environments in amorphous solids. It is an important calculation 
to verify  credibility of a model.\cite{Lopinski1} Meanwhile, it is equally challenging to get a good comparison of vibrational properties between theoretical and 
experimental results. A lot of factors like: model size, completeness of basis set, etc. can affect vibrational properties. We have performed ionic-relaxation on our models to attain a local minimum with forces on each atom less than ($\sim$ 0.01 eV/atom) while simultaneously relaxing 
lattice vectors to zero pressure. This results in slightly different number density and a non-orthogonal cell but as 
shown in our earlier work,\cite{Bhattarai2} it is crucial to have coordinates well relaxed before evaluating vibrational properties of the models.

We have computed vibrational properties for our four models(216 MQ, 216 FEAR, 512 FEAR and 1024 FEAR) using the dynamical matrix. We displaced each atom in 6-directions($\pm x$,$\pm y$,$\pm z$) with a small displacement of ($\sim$ 0.015 $\mathring{A}$). After, each of these small displacement an \textit{ab initio} force calculation was carried out to obtain force 
constant matrix (see details \cite{Bhattarai1}). The VDOS for amorphous systems with \textit{N} number of atoms is defined as,

\begin{equation}
g(\omega)=\frac{1}{3N} \sum_{i=1}^{3N} \delta(\omega-\omega_i)
\end{equation}

We have computed the VDOS for our models using the method of Gaussian broadening with a standard deviation of $\sigma=1.86$ meV or 15.0 $cm^{-1}$.
The first three zero frequency modes are due to supercell translations, and have been neglected during our calculations of VDOS and vibrational IPR.
We report the VDOS for our different models in Fig. 6.

As seen in Fig. 6, there is a slight horizontal shift in VDOS depending upon system size and completeness of basis set. VDOS calculated with minimal basis
set (single-$\zeta$, SZ) in SIESTA has a qualitative agreement with the experimental result, while slight shift is observed at both low and high energies w.r.t 
the experiment. This result is refined by using a more complete basis-set (double-$\zeta$, DZ), which gives us a better agreement of our models with the experiment.
We have computed VDOS using DZ for two of our models (FEAR 216 and FEAR 512). The VDOS obtained for FEAR 512 is strikingly similar to the experiment (Fig.6, right panel).
This switch from minimal basis to double $\zeta$ basis impacts computation time needed for these calculations and with our resources in hand we 
simply could not perform a DZ calculations for our FEAR 1024 atom system.

Thus, we can infer completeness of basis-set 
affects these low energy excitation of atoms in amorphous silicon. The most remarkable feature is the improvement at high frequencies. 
Based on our zero pressure (double-$\zeta$, DZ)
calculation, it's agreement with experimental VDOS and specific heat (Fig. 7), we predict new density for \asi. Our predicted results are 
tabulated in Table. II and our results for the zero pressure (double-$\zeta$, DZ) calculation is close to the experimentally predicted
density for \asi (2.28 $g/cm^3$).\cite{Custer1}

\begin{table}[!ht]
\caption{\label{tab:table3}Details of densities obtained after zeropressure relaxation of FEAR models for single-$\zeta$ (SZ) and double-$\zeta$(DZ) basis
sets in SIESTA. Our density for zero pressure (DZ) is closer to the experimental density\cite{Custer1} at 2.28 $g/cm^3$.}
\setlength{\arrayrulewidth}{0.50mm}
\vspace{0.3cm}
\begin{tabular}{|c|c|c|c|}
\hline
\rule{0pt}{4mm}
 
 Models & Volume($\AA^3$)& N($atom/\AA^3$)  & $\rho(g/cm^3)$\\ \hline
\rule{0pt}{4mm}

216 FEAR(SZ) & 4643.77 & 0.046514 & 2.16   \\
\rule{0pt}{4 mm}

512 FEAR(SZ) & 10997.33 & 0.046557 & 2.17          \\
\rule{0pt}{4 mm} 

1024 FEAR(SZ) & 21755.17 & 0.047067 & 2.19           \\  % inserting body of the table
 \hline \hline
 \rule{0pt}{4 mm}
 
216 FEAR (DZ) & 4510.57 & 0.047887 & 2.23            \\
\rule{0pt}{4 mm}

512 FEAR(DZ) & 10652.76 &0.048062& 2.24              \\
\rule{0pt}{4 mm}

1024 FEAR(DZ) & 21213.92 & 0.048270 & 2.25               \\ 
\hline
\end{tabular}

\end{table}

Structural disorder in amorphous solids leads to localized modes and these localized modes can be evaluated by defining a quantity, the inverse participation 
ratio (IPR). Similar to electronic IPR, we can evaluate vibrational IPR using the obtained normalized displacement vectors. The IPR can be readily evaluated with the 
obtained normalized displacement vectors ($u^j_i$), $\mathcal{I}$ for the vibrations can be defined as (for $j^{th}$ mode), 

\begin{equation}
\mathcal{I}=\frac{\sum_{i=1} ^N |u^j_i|^4 }{\big(\sum_{i=1} ^N |u^j_i|^2\big )^2}
\end{equation}

The inverse participation ratio value of a localized mode is $ \approx 1 $ and for an extended mode is almost equal to zero. We have plotted IPR of our four 
models in Fig.6 (right panel). The vibrations at low energies are mostly extended modes, these represent mostly bending type while vibrations at higher 
energies are dominated by stretching type of modes.\cite{Bhattarai1,Bhattarai2}

\subsubsection{Specific Heat in the harmonic approximation}

\begin{figure}[!ht]
  \centering
  \includegraphics[width=0.95\linewidth]{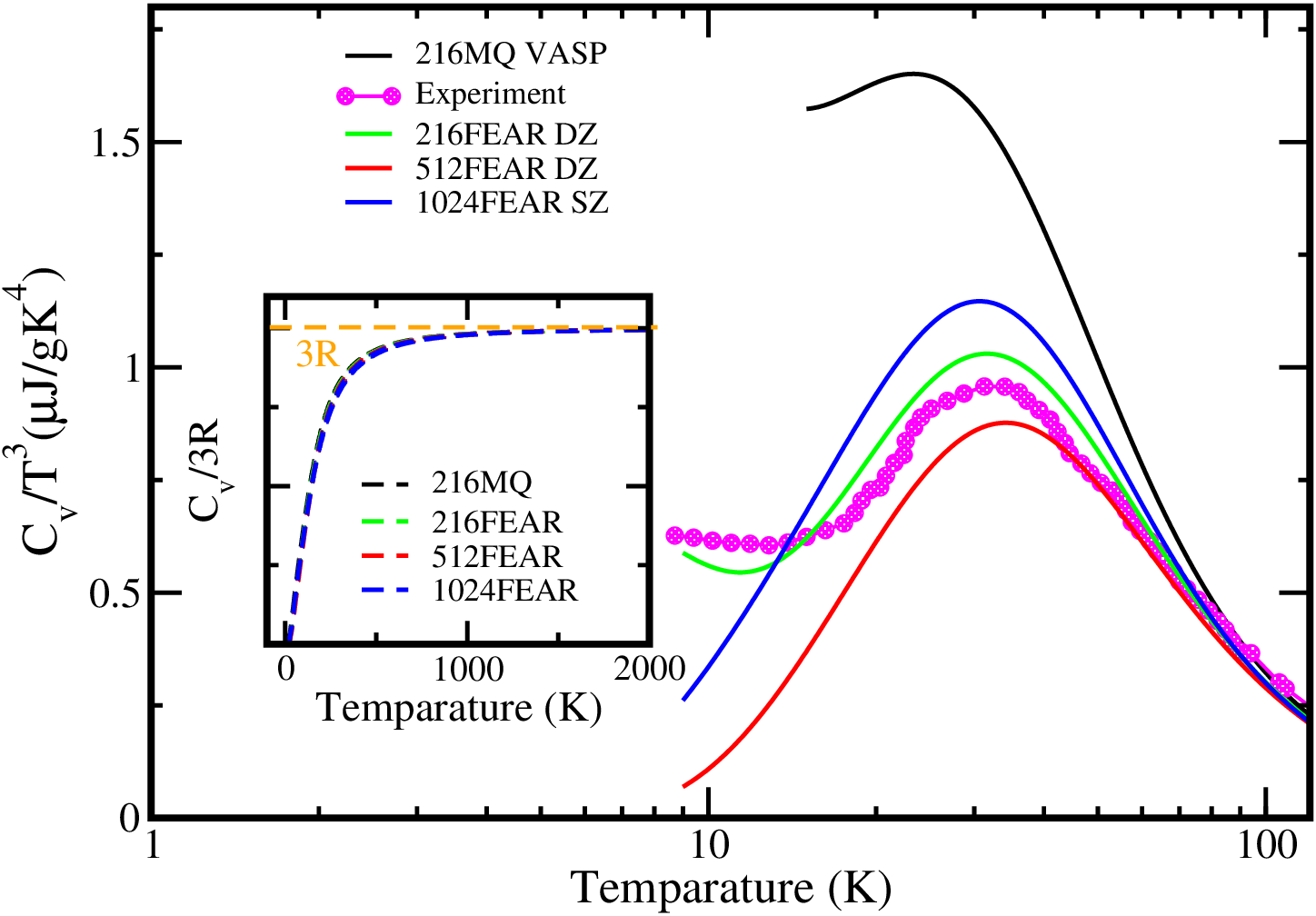}
\caption{(Color online) Plot of specific-heat ($C_v/T^3$) for the four models compared with the experimental results\cite{Zink1}. The inset shows the classical
(\textit{Dulong-petit}) limit at higher temperature.}
  
\label{fig:fig102466789}
\end{figure}

We evaluate the specific heat in the harmonic approximation using information of vibrational density of states~$g(\omega)$ obtained for our models.  We compute the specific heat $C_v(T)$ from the relation\cite{Maradudin1}
\begin{equation}
 C(T)= 3R\int_0^{E_{max}}\Bigg(\frac{E}{k_{B}T}\Bigg)^2 \frac{e^{E/k_BT}}{\Big(e^{Ek_BT}-1\Big)^2}g(E)dE
\end{equation}
Here, the $g(E)$ is normalized to unity\cite{Bhattarai2,Nakhmanson1}. Our plot for specific heat is shown in Fig. 7. We have a qualitative agreement with 
the experiment for our four models while the peak around ($\sim 30 K$) is largely affected by the quality of VDOS obtained. Our three models FEAR-216(DZ),
FEAR-512(DZ) and FEAR-1024(SZ) improves the previously agreement of different models with the experiment.\cite{Zink1} 

We infer from our calculation of VDOS and specific heat that a bigger size model together with a bigger basis set gives us a better understanding 
of these low energy excitations. This further outlines the importance of our method FEAR, with the resources available to us it is not possible to 
fabricate melt-quench models of size 512 and 1024 atoms.

\section{\label{Comparison} Conclusions} 

This paper presents an investigation pertaining to the complex amorphous material (\asi), which was evaluated with respect to its
structural, electronic and vibrational properties. Various model types, MQ and FEAR, were constructed of different sizes for this investigation.
Our results reveal that the recently developed FEAR method provides an accurate outcome, which correlates quite well with experimental data, 
even for relatively large structures sizes (512 and 1024). To our knowledge our VDOS result depicts the most clear picture of low energies excitations
for \asi. We also 
predict new density of amorphous silicon based on \textit{ab initio} minimum, our prediction is remarkably close to the experimentally found density.

\section{\label{Acknowledgment} Acknowledgment} 
The authors are thankful to the NSF under grant numbers DMR 1506836, DMR 1507118 and DMR 1507670.
We would like to thank Dr. Anup Pandey for helpful conversations. We also are thankful for the financial support from 
Condensed Matter and Surface Science (CMSS) at Ohio University.
Lastly, we acknowledge computing time provided by the Ohio Supercomputer Center for this research. We also thank NVIDIA Corporation for donating a Tesla K40 GPU which was used in some of these computations. \\

\bibliography{sample}

\end{document}